\RequirePackage[2020-02-02]{latexrelease}
\documentclass[twocolumn,prc,superscriptaddress,amsmath,amssymb]{revtex4-2}
\usepackage{graphicx}% Include figure files
\usepackage{dcolumn}% Align table columns on decimal point
\usepackage{bm}% bold math
\usepackage{color}
\usepackage{float}
\usepackage{amsmath}

\usepackage[colorlinks,
           bookmarks=true,
           linkcolor=blue,
           urlcolor=blue,
            anchorcolor=black,
            citecolor=blue
            ]{hyperref}

\usepackage[all]{hypcap}

\begin{document}

\title{Intermittency analysis of proton numbers in heavy-ion collisions at energies available at the BNL Relativistic Heavy Ion Collider}
\author{Jin Wu}
\affiliation{Key Laboratory of Quark and Lepton Physics (MOE) and Institute of Particle Physics, \\
Central China Normal University, Wuhan 430079, China}
\author{Yufu Lin}
\affiliation{Key Laboratory of Quark and Lepton Physics (MOE) and Institute of Particle Physics, \\
Central China Normal University, Wuhan 430079, China}
\author{Zhiming Li}
\email{lizm@mail.ccnu.edu.cn}
\affiliation{Key Laboratory of Quark and Lepton Physics (MOE) and Institute of Particle Physics, \\
Central China Normal University, Wuhan 430079, China}
\author{Xiaofeng Luo}
\email{xfluo@mail.ccnu.edu.cn}
\affiliation{Key Laboratory of Quark and Lepton Physics (MOE) and Institute of Particle Physics, \\
Central China Normal University, Wuhan 430079, China}
\author{Yuanfang Wu}
\email{wuyf@mail.ccnu.edu.cn}
\affiliation{Key Laboratory of Quark and Lepton Physics (MOE) and Institute of Particle Physics, \\
Central China Normal University, Wuhan 430079, China}

\begin{abstract}
Local density fluctuations near the QCD critical point has been suggested to exhibit a power-law behavior which can be probed by an intermittency analysis on scaled factorial moment (SFM) in relativistic heavy-ion collisions. The collision energy and centrality dependence of the second-order SFMs are systematically investigated in Au $+$ Au collisions at $\sqrt{s_{NN}}$ = 7.7, 11.5, 19.6, 27, 39, 62.4, and 200 GeV within the UrQMD model. We estimate the noncritical background in the measurement of intermittency and propose a cumulative variable method to effectively remove the background contributions. We further study the effect of particle detection efficiency by implementing the RHIC (STAR) experimental tracking efficiencies in the UrQMD events. A cell-by-cell method is proposed for experimental application of efficiency corrections on SFM. This work can provide a guidance of background subtraction and efficiency correction for the experimental measurement of intermittency in the search of the QCD critical point in heavy-ion collisions.
\end{abstract}
\maketitle
%%%%%%%%%%%%%%%%%%%%%%%%%%%%%%%%%%%%%%%%%%%%%%%%%%%%%%%%%%%%%%%%%%%%%%%%%%%%%%%
\section{INTRODUCTION}
One of the major goals of relativistic heavy-ion collisions is to explore the QCD phase structure~\cite{StephanovPD,MomentOverview,QCDReport,conservecharge0,conservecharge1}. Theoretical studies have shown the existence of a critical point (CP) at finite baryon chemical potential and temperature~\cite{CEP1, CEP2, CEP3}. This CP is proposed to be characterized by a second-order phase transition, which becomes a unique property of strongly interacting matter~\cite{searchCEP1,searchCEP2,searchCEP3,searchCEP4,STARPRLMoment}. In the thermodynamic limit, the correlation length diverges at the CP and the system becomes scale invariant and fractal~\cite{invariant1,invariant2,invariant3}. It has been shown that the density fluctuations near the QCD critical point form a distinct pattern of power-law or intermittency behavior in the matter produced in high energy heavy-ion collisions~\cite{Antoniou2006PRL,Antoniou2010PRC,Antoniou2016PRC,Antoniou2018PRD}.

Intermittency is a manifestation of the scale invariance, fractality of  physical processes and the stochastic nature of the underlying scaling law~\cite{invariant3}. It can be revealed in transverse momentum spectra as a power-law behavior of scaled factorial moment (SFM)~\cite{invariant1, Antoniou2006PRL}. In current high-energy heavy-ion experiments at CERN SPS~\cite{NA49SFM,NA61universe}, the NA49 and NA61 collaborations have performed the intermittency analysis with various sizes of colliding nuclei. A power-law behavior has been observed in Si + Si collisions at 158A GeV~\cite{NA49SFM}. Recently, the STAR collaboration reported the preliminary result of charged-particle intermittency from the beam energy scan (BES) program at RHIC. The critical exponent extracted from intermittency index shows a minimum in central Au + Au collisions around $\sqrt{s_\mathrm{NN}}$ = 20-30 GeV ~\cite{STARintermittency}. In the mean time, various model studies have been conducted to investigate the unique behavior of intermittency under various underlying mechanisms~\cite{CMCPLB,UrQMDLi,IJMPE,amptNPA,epos}.

For a self-similar system with intermittency, it is expected that the multiplicity distribution in momentum space is associated with a strong clustering effect, which indicates a remarkably structured phase-space density~\cite{invariant3,densityF}. However, the inclusive single-particle multiplicity spectra in finite space of high-energy collisions are significantly influenced by background effects. The multiplicity distribution is constrained or modified by conservation law, resonance decay, statistical fluctuations, etc. It has been shown that the statistical fluctuations due to a finite number of particles~\cite{SFM1} or the choices of the size in momentum space~\cite{APPB} will influence the measured SFM. Therefore, it is necessary to estimate and remove these trivial effects in order to get a clean power-law exponent, and then one can compare the measured intermittency with theoretical predictions. For this purpose, Ochs~\cite{OchsCumulative}, Bialas and Gazdzicki~\cite{BialasCumulative} proposed to study intermittency by using the cumulative variable method, in which the single-particle density is a constant. The cumulative variable method can effectively reduce the distortions of the simple scaling law caused by a nonuniform single-particle spectrum and there is no bias from the shape of the inclusive distribution. We will study how to remove the trivial background effects by the cumulative variable method in the measurement of SFMs in heavy-ion collisions.

In heavy-ion collision experiments, the particle detector has a finite detection efficiency, which could simply result from the limited capability of the detector to register the finite-state particles~\cite{STAREfficiency,EfficiencyLuo}. This will lead to the loss of particle multiplicity in an event, which makes the measured event-by-event multiplicity distributions in momentum space different from the original produced ones~\cite{EfficiencyLuo,MomentOverview}. The values of SFM could be significantly modified by the detector efficiency, which will distort the original signal possibly induced by the CP.  Therefore, we should recover the SFM of the true multiplicity distributions from the experimentally measured ones by applying proper efficiency correction technique. 

The paper is organized as follows. A brief introduction to the UrQMD model is given in Sec. II. In sec. III, we introduce the method of intermittency analysis by using SFMs. Then, the collision energy and centrality dependence of SFMs are investigated by UrQMD model in Au $+$ Au collisions from $\sqrt{s_\mathrm{NN}}$ = 7.7 to 200 GeV. In Sec. V, we discuss the estimation and subtraction of background in the calculations of SFMs. In Sec. VI, the efficiency correction formula is deduced, followed by a check of the validity of the method by the UrQMD model. Finally, we give a summary and outlook of this work.

%%%%%%%%%%%%%%%%%%%%%%%%%%%%%%%%%%%%%%%%%%%%%%%%%%%%%%%%%%%%%%%%%%%%%%%%%%%%%%%     
\section{ULTRA RELATIVISTIC QUANTUM MOLECULAR DYNAMICS MODEL}                               
The ultra relativistic quantum molecular dynamics (UrMQD) model is a microscopic many-body model that has been extensively applied to simulate $p + p$, $p + A$, and $A + A$ interactions in ultrarelativistic heavy-ion collisions~\cite{MBUrQMD,SABassUrQMD,HPUrQMD}. It provides phase-space descriptions of different reaction mechanisms based on the covariant propagation of all hadrons with stochastic binary scattering, color string formation and resonance decay~\cite{SABassUrQMD}. This model includes baryon-baryon, meson-baryon and meson-meson interactions with more than 50 baryon and 45 meson species. It preserves the conservation of electric charge, baryon number, and strangeness number. It models the phenomena of baryon stopping which is an essential feature encountered in heavy ions at low beam energies. It is a well-designed transport model~\cite{MBUrQMD} for simulations with the entire available range of energies from SIS energy ($\sqrt{s_\mathrm{NN}}$ = 2 GeV) to the top RHIC energy ($\sqrt{s_\mathrm{NN}}$ = 200 GeV). More details about the UrQMD model can be found in Refs.~\cite{MBUrQMD,SABassUrQMD,HPUrQMD}.

The UrQMD model is a suitable simulator to estimate the noncritical contribution and other trivial background effects in the measurement of correlations and fluctuations in heavy-ion collisions. In this work, we use the UrQMD model (version 2.3) to generate Monte Carlo event samples of Au $+$ Au collisions at RHIC energies. The corresponding event statistics are 72.5, 105, 106, 81, 133, 38, 56 millions at $\sqrt{s_\mathrm{NN}}$ = 7.7, 11.5, 19.6, 27, 39, 62.4, 200 GeV, respectively.

%-------------Figure about energy and centrality dependence of FM
 \begin{figure*}
     \centering
     \includegraphics[scale=0.85]{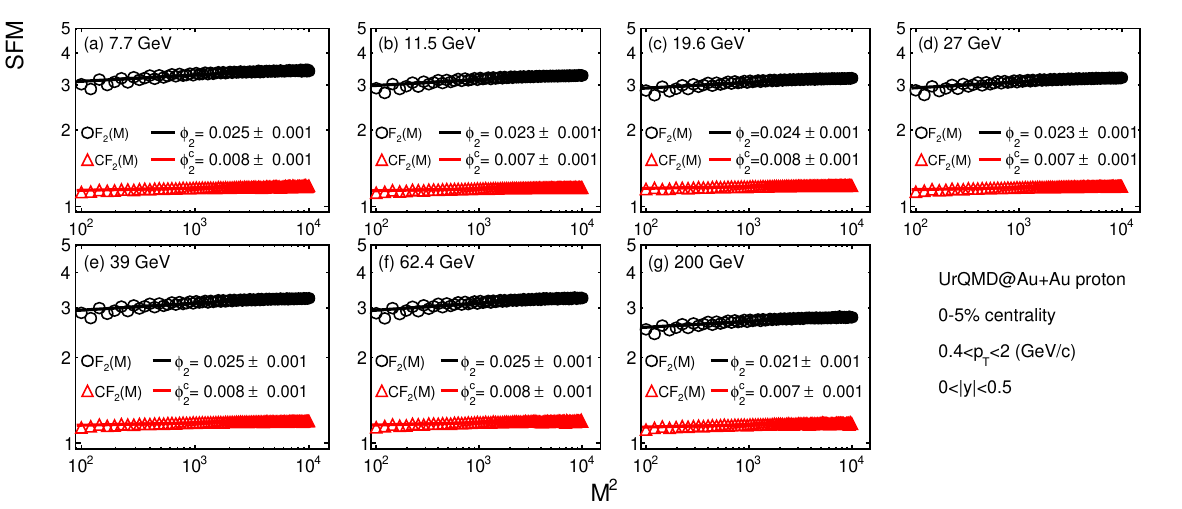}
      \caption{The second-order scaled factorial moment (black circles) $F_{2}(M)$ as a function of number of partitioned cells in a double-logarithmic scale at $\sqrt{s_\mathrm{NN}}$ = 7.7-200 GeV from the UrQMD model. The black lines are the power-law fitting. The corresponding red ones represent the SFMs calculated by the cumulative variable method.}
     \label{Fig:F2E}
     \end{figure*}
 
     \begin{figure*}
     \centering 
     \includegraphics[scale=0.85]{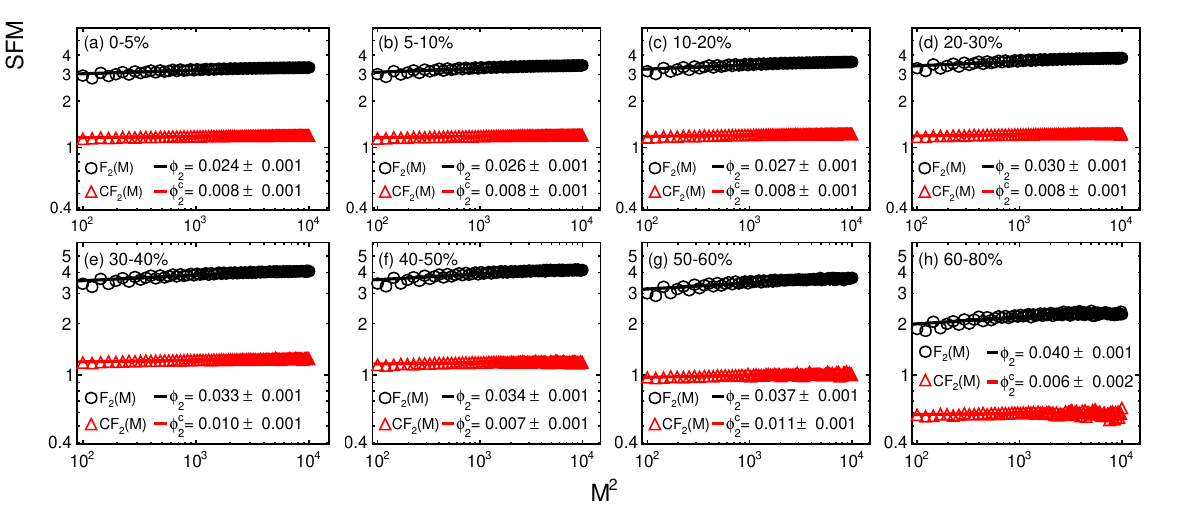}
      \caption{The second-order scaled factorial moment (black circles) as a function of number of partitioned cells from the most central ($0\%-5\%$) to the most peripheral ($60\%-80\%$) collisions at $\sqrt{s_\mathrm{NN}}$ = 19.6 GeV. The corresponding red ones represent the SFMs calculated by the cumulative variable method.}
     \label{Fig:F2C}
     \end{figure*}
     
%%%%%%%%%%%%%%%%%%%%%%%%%%%%%%%%%%%%%%%%%%%%%%%%%%%%%%%%%%%%%%%%%%%%%%%%%%%%%%%
\section{Method of Analysis}
It is argued that in heavy-ion collisions large baryon density fluctuations may provide a unique signal to the phase transition in the QCD phase diagram. It is expected to observe the critical density fluctuations as a power-law pattern on available phase-space resolution if the system freezes out right in the vicinity of the critical point~\cite{Antoniou2006PRL,NA61universe}.

%In high-energy experiments, the power-law or intermittency behavior can be measured by calculations of SFMs of baryon number density~\cite{HattaPRL,Antoniou2006PRL, NA61universe}. For this purpose, an available region of momentum space is partitioned into $M^{D}$ equal-size bins. The $q$th order SFM, i.e., $F_{q}(M)$, is defined as\\
\begin{equation}
F_{q}(M)=\frac{\langle\frac{1}{M^{D}}\sum_{i=1}^{M^{D}}n_{i}(n_{i}-1)\cdots(n_{i}-q+1)\rangle}{\langle\frac{1}{M^{D}}\sum_{i=1}^{M^{D}}n_{i}\rangle^{q}},
 \label{Eq:FM}
\end{equation}

\noindent with $M^{D}$ being the number of cells in the $D-$dimensional partitioned momentum space, $n_{i}$ the measured multiplicity in the $i$th cell, and $q$ the order of the moment.

%If the system exhibits critical fluctuations, SFM is expected to follow a scaling function:~\cite{Antoniou2006PRL,NA49SFM}\\
\begin{equation}
F_{q}(M)\sim (M^{D})^{\phi_{q}}, M\rightarrow\infty.
 \label{Eq:PowerLaw}
\end{equation}

\noindent A power-law behavior of $F_{q}(M)$ on the partitioned number $M^{D}$ when $M$ is large enough is referred to as intermittency. $M$ is the number of cells in each dimension. The scaling exponent $\phi_{q}$ is called the intermittency index that characterizes the strength of the intermittency behavior. By using a critical equation of state of a three-dimensional (3D) Ising system, the second-order intermittency index in a two-dimensional transverse momentum space is predicted to be $\phi_{2}=\frac{5}{6}$~\cite{Antoniou2006PRL} for baryon density and $\phi_{2}=\frac{2}{3}$ for sigma condensate~\cite{NGNPA2005}. The search of multiplicity fluctuations in increasing number of partition intervals using the SFM method was first proposed in Refs.~\cite{SFM1,SFM2}. Recent studies show that one can probe QCD critical fluctuations~\cite{CMCPLB} and estimate the possible critical region~\cite{Antoniou2018PRD} from intermittency analysis in relativistic heavy-ion collisions.

In the following section, we calculate the second-order SFM of proton density in transverse momentum space by using event samples from the UrQMD model in Au $+$ Au collisions at $\sqrt{s_\mathrm{NN}}$ = 7.7-200 GeV. Then, the intermittency index, $\phi_{2}$, can be extracted by fitting Eq.~\eqref{Eq:PowerLaw}.

 %%%%%%%%%%%%%%%%%%%%%%%%%%%%%%%%%%%%%%%%%%%%%%%%%%%%%%%%%%%%%%%%%%%%%%%%%%%%%%%     
\section{Energy and centrality dependence of Scaled Factorial moments}
By using the UrQMD model, we generate event samples at various centralities in Au $+$ Au collisions at $\sqrt{s_\mathrm{NN}}$ = 7.7, 11.5, 19.6, 27, 39, 62.4, and 200 GeV. In the model calculations, we apply the same kinematic cuts and technical analysis methods as those used in the RHIC (STAR) experiment data~\cite{net_proton2014}. The protons are measured at midrapidity($|y|<0.5$) within the transverse momentum $0.4<p_{T}<2.0$ GeV/c. The centrality is defined by the charged pion and kaon multiplicities within pseudorapidity $|\eta|<1.0$. Since we only concern protons in the calculations and use pions and kaons without protons to determine centrality, it can effectively avoid auto-correlation effects in the measurement of SFMs. In our analysis, we focus on proton multiplicities in a two-dimensional transverse momentum space of $p_{x}$ and $p_{y}$. The available two-dimensional (2D) region of transverse momentum is partitioned into $M^2$ equal-size bins to calculate SFMs in various sizes of cells. The statistical error is estimated by using the Bootstrap method~\cite{BEboostrap}.

In Fig. 1, the black circles represent the second-order SFMs as a function of number of partitioned bins, directly calculated in transverse momenta for proton numbers in $0\%-5\%$ the most central Au $+$ Au collisions at $\sqrt{s_\mathrm{NN}}$ = 7.7-200 GeV. It is observed that $F_{2}(M)$ increases slowly with increasing number of dividing bins. The black lines show the power-law fit of $F_{2}(M)$ according to Eq.~\eqref{Eq:PowerLaw}. The slopes of the fitting, i.e., the intermittency indices $\phi_{2}$, are found to be small at all energies. And they are much less than the theoretical prediction $\phi_{2}=5/6$ for a critical system of the 3D Ising universality class~\cite{Antoniou2006PRL}.

The $F_{2}(M)$ measured at various collision centralities in Au $+$ Au collisions at $\sqrt{s_\mathrm{NN}}$ = 19.6 GeV are shown as the black circles in Fig. 2. And the black lines are the fitting according to Eq.~\eqref{Eq:PowerLaw}. Again, we find that the directly calculated SFMs can be fitted with a small intermittency index. The values of $\phi_{2}$ increase slightly from the most central ($0\%-5\%$) to the most peripheral ($60\%-80\%$) collisions.

Therefore, we observe that the intermittency indices from the directly calculated SFMs are small but nonzero in Au $+$ Au collisions from the UrQMD model. However, the UrQMD model~\cite{MBUrQMD} is a transport model which does not include any critical related self-similar fluctuation. In this case, the SFM will be independent of the number of partition bins and $\phi_{2}$ would be expected to be zero~\cite{invariant3,SFM1,SFM2}. So there must exist some trivial noncritical contributions from background. The similar small values of intermittency indices are also found in the HIJING~\cite{NA49C+C} and PYTHIA~\cite{pythia} Monte Carlo models. In the NA49 and NA61 experimental data, the observed scaling behavior can be reproduced by mixing critical events with a probability of more than 90$\%$ with uncorrelated random tracks~\cite{NA49SFM,NA61Ar+Sc}. In the case that critical related tracks are rare, the underlying critical signal may be diluted by majority random tracks. In light of such a scenario, precise knowledge of the background is crucial to the measurement of intermittency in heavy-ion collisions. We will investigate how to remove the background effects from the directly measured SFMs in the next section.

%%%%%%%%%%%%%%%%%%%%%%%%%%%%%%%%%%%%%%%%%%%%%%%%%%%%%%%%%%%%%%%%%%%%%%%%%%%%%%% 
\section{Background Subtraction}
%%%%%%%%%%%%%%%%%%%%%%%%%%%%%%%%%%%%%%%%%%%%%%

To extract the signature of critical fluctuations, it is essential to understand the noncritical effects or background contributions on the experimental observables. The background effects will change the multiplicity distributions in the measured finite momentum space. Then the multiplicity in each partitioned cell, $n_i$, will be modified accordingly when calculating SFMs based on Eq.~\eqref{Eq:FM}. Since the values of calculated SFMs are changed, the intermittency index will be affected. It is shown that the SFMs are significantly influenced when adding uncorrelated particles from background to the event samples of self-similar signals~\cite{resolution}. In this purpose, NA49 and NA61 use the mixed event method to estimate and subtract background by assuming that the particle multiplicity in each cell can be simply divided into background and critical contributions~\cite{NA49SFM}. In this paper, we pursue the cumulative variable method, which has been proved to drastically reduce distortions of intermittency due to nonuniform single-particle density from background contributions~\cite{OchsCumulative,BialasCumulative,na22}, to understand and remove the background effects.

The cumulative variable $X(x)$ is related to the single-particle density distribution $\rho(x)$ through~\cite{BialasCumulative,OchsCumulative}:

\begin{equation}
  \large X(x)=\frac{\int_{x_{min}}^{x} \rho(x)dx}{\int_{x_{min}}^{x_{max}}\rho(x)dx}.
 \label{Eq:cvariable}
\end{equation}

\noindent Here $x$ represents the original measured variable, e.g., $p_{x}$ or $p_{y}$. $\rho(x)$ is the density function of $x$. $x_{min}$ and $x_{max}$ are the lower and upper phase-space limits of the chosen variable $x$.

The cumulative variable $X(x)$ is determined by the shape of density distribution $\rho(x)$. The distribution of the new variable $X(x)$ is uniform in the interval from 0 to 1. It has been proved that the cumulative variable could remove the dependence of the intermittency parameters on the shape of particle density distributions and give a new way to compare measurements from different experiments~\cite{BialasCumulative}. To use the cumulative variable, the two-dimensional momentum space $p_{x}p_{y}$ which is partitioned into $M^{2}$ cells will transfer to be $p_{X}p_{Y}$ space. And the SFM directly calculated in $p_{x}p_{y}$ space [$F_{2}(M)$] will transfer to be $CF_{2}(M)$, which is now calculated in $p_{X}p_{Y}$ space. The process of fitting $\phi_{2}^{c}$ from $CF_{2}(M)$ is similar to that of $\phi_{2}$ from $F_{2}(M)$ according to Eq.~\eqref{Eq:PowerLaw}.

%%%%%%%%%%%%%%%%%Fig cumulative aviable
\begin{figure}[htp]
     \centering
     \includegraphics[scale=0.45]{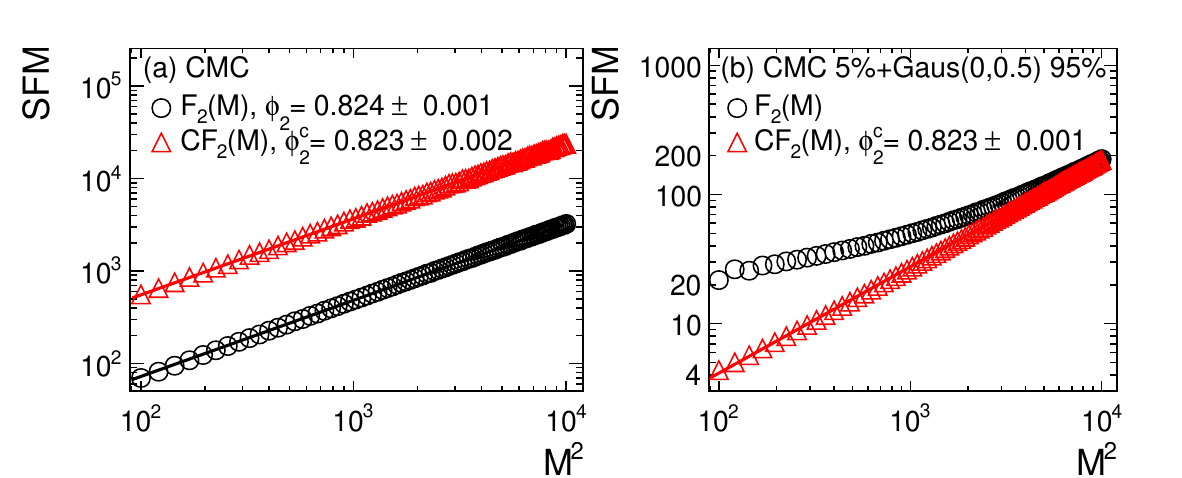}
     \label{Fig:GasandCMC}
     \caption{The black symbols represent the second-order scaled factorial moment as a function of number of partitioned cells (a) in pure CMC events and (b) in CMC events contaminated with Gaussian background fluctuations. The corresponding red ones are the SFMs calculated by the cumulative variable method.}
\end{figure}
To test the validity of the cumulative variable method in the calculations of SFMs, we use a critical Monte Carlo (CMC) model~\cite{Antoniou2006PRL,CMCPLB} of the 3D Ising universality class to generate critical event samples. The CMC model involves the self-similar or intermittency nature of particle correlations and leads to an intermittency index of $\phi_{2}=\frac{5}{6}$~\cite{Antoniou2006PRL}. In Fig. 3(a), both $F_{2}(M)$ (black circles) and $CF_{2}(M)$ (red triangles) are shown in the same panel. We observe that $CF_{2}(M)$ follows a good power-law behavior as $F_{2}(M)$ with increasing $M^{2}$. Within statistical errors, the intermittency index $\phi_{2}^{c}$ fitted from $CF_{2}(M)$ equals $\phi_{2}$ obtained from $F_{2}(M)$. It means that the cumulative variable method does not change the intermittency behavior for a pure critical signal event sample, which has been proved by Bialas and Gazdzicki when they proposed to use the cumulative variable method to study intermittency~\cite{BialasCumulative}. In Fig. 3(b), the CMC event sample is contaminated by hand with a statistical Gaussian background contribution, with the mixed probability $\lambda = 95\%$. The chosen value of $\lambda$ is close to the one used in the simulations of background in the NA49 experiment~\cite{NA49SFM}. In this case, one finds that the directly calculated $F_{2}(M)$ deviates substantially from the linear dependence, i.e., violation of the scaling law because of the Gaussian background contribution. So we can not make a good fitting based on the scaling function defined in Eq.~\eqref{Eq:PowerLaw}. However, the trend of $CF_{2}(M)$, which is calculated by the cumulative variable method, still obeys a similar power-law dependence on $M^{2}$ as that in Fig. 3(a). Furthermore, the intermittency index $\phi_{2}^{c}$ calculated from $CF_{2}(M)$ keeps unchanged when comparing to the one in original CMC sample shown in Fig. 3(a). We feel that these results are encouraging. They confirms that, in the intermittency analysis, the cumulative variable method efficiently removes the effects caused by the background contribution.

Let us go back to the problem of the background effect in the UrQMD model from in Sec. IV. We calculate SFMs in the same event sample by the proposed cumulative variable method and then get the intermittency index from $CF_{2}(M)$. The results are shown as red triangles and red lines in Fig. 1 and Fig. 2. $CF_{2}(M)$ is found to be nearly flat with an increasing number of cells in all measured energies and centralities. Furthermore, the intermittency index, with the value near to zero, is much smaller than the value directly calculated from $F_{2}(M)$. It verifies that the background of noncritical effect can be efficiently removed by the cumulative variable method in the calculation of SFMs in the UrQMD model. This method could also be used for the intermittency analysis in the ongoing experimental at RHIC (STAR) or further heavy-ion experiments in search of the QCD critical point. We would also note that the fit values of $\phi_{2}^{c}$ from $CF_{2}(M)$ are still not exactly zero although they are much smaller than  $\phi_{2}$ obtained directly from measured $F_{2}(M)$. It possibly accounts for other effects such as proton correlations due to Coulomb repulsion and Fermi–Dirac statistics~\cite{NA49SFM} or the influence of momentum resolution~\cite{resolution}. Further studies on these effects should also be concerned in the calculation of intermittency index in heavy-ion collisions. 

%%%%%%%%%%%%%%%%%%%%%%%%%%%%%%%%%%%%%%%%%%%%%%%%%%%%%%%%%%%%%%%%%%%%%%%%%%%%%%%
\section{Efficiency correction}
One of the difficulties of measuring SFMs and intermittency in experiment is efficiency correction. It is known that the values of SFMs are changed from the original true SFMs due to the fact that detectors miss some particles with a probability named efficiency. To understand the underlying physics associated with this measurement, one needs to perform a careful study on the efficiency effect. Generally, the efficiencies in experiments are obtained by using Monte Carlo (MC) embedding technique~\cite{STAREfficiency,embedding}. This allows for the determination of the efficiency, which is the ratio of the matched MC tracks number and the number of input tracks. It contains the effects of tracking efficiency, detector acceptance and interaction losses. 

Let us denote the number of produced particles as $N$ and the number of experimental measured ones as $n$ with a detection efficiency $\epsilon$. To correct the factorial moment for efficiency effects, one has to invoke a model assumption for the response of the detector. It is often assumed to follow a binomial probability distribution function~\cite{EfficiencyMasakiyo, EfficiencyPremomoy, EfficiencyAdam, EfficiencyKoch}. Then the probability to measure $n$ particles given $N$ produced particles can be expressed as
\begin{equation}
p(n|N)=B(n,N;\epsilon)=\frac{N!}{n!(N-n)!}\epsilon^{n}(1-\epsilon)^{N-n}.
\label{Eq:binomial}
\end{equation}

The true factorial moment is defined as $f_{q}^{true}=\langle N(N-1)...(N-q+1)\rangle$. It can be recovered by dividing the measured factorial moment, $f_{q}^{measured}=\langle n(n-1)...(n-q+1)\rangle$, with appropriate powers of the detection efficiency~\cite{EfficiencyLuo,EfficiencyAdam, EfficiencyKoch,EfficiencyHeshu, momentHADES}: 
\begin{equation}
 f_{q}^{corrected}=\frac{f_{q}^{measured}}{\epsilon^{q}}=\frac{\langle n(n-1)...(n-q+1)\rangle}{\epsilon^{q}}.
 \label{Eq:f2correction}
\end{equation}

\noindent This strategy has been used for the efficiency corrections in the high-order cumulant analysis~\cite{EfficiencyLuo,MomentOverview, EfficiencyFM, EfficiencyToshihiro, EfficiencyKoch,STARMoment}. Consider that the probability to detect a particle is governed by a binomial distribution, then both cumulants~\cite{EfficiencyLuo,MomentOverview} and off-diagonal cumulants~\cite{EfficiencyKoch} can be expressed in term of factorial moments and then can be corrected by using Eq.~\eqref{Eq:f2correction}.

We apply the strategy for the efficiency correction to SFMs defined in Eq.~\eqref{Eq:FM}. Since the available region of phase  space is partitioned into a lattice of $M^{2}$ equal-size cells, every element $\langle n_{i}(n_{i}-1)...(n_{i}-q+1)\rangle$ of measured SFMs should be corrected one by one. In this way, the efficiency corrected SFM is deduced as
 
\begin{eqnarray}
  F_{q}^{corrected}(M)=\frac{\langle\frac{1}{M^{2}}\sum_{i=1}^{M^{2}}\frac{n_{i}(n_{i}-1)\cdots(n_{i}-q+1)}{\bar{\epsilon_{i}}^{q}}\rangle}{\langle\frac{1}{M^{2}}\sum_{i=1}^{M^{2}}\frac{n_{i}}{\bar{\epsilon_{i}}}\rangle^{q}}.
 \label{Eq:FMcorrection}
 \end{eqnarray}

\noindent Here, $n_{i}$ denotes the number of measured particles located in the $i$-th cell. The mean $\bar{\epsilon_{i}}$, is calculated by $\langle\frac{\sum_{j=1}^{n_i}\epsilon_i^j}{n_i}\rangle$, representing the event average of the mean efficiency for the particles located in the $i$th cell. Its value depends on the momentum range of the $i$-th cell and particle species in experimental measurement~\cite{momentHADES,STARMoment}. We may note that the efficiencies are assumed to be uncorrelated between cells. The possible dynamical fluctuations in the efficiencies due to time and/or detector variations should be handled in experiments. We call the efficiency correction technique of Eq.~\eqref{Eq:FMcorrection} the cell-by-cell method.

To demonstrate the validity of the cell-by-cell method, we employ the UrQMD model with the particle detection efficiencies used in real experiments. It is simulated by injecting particle tracks from UrQMD events into the RHIC (STAR) detector acceptance with the experimental efficiencies. In the STAR experiment, the detection efficiency is not a constant but depends on the momentum of particles~\cite{STAREfficiency,STARMoment,LuoCPOD}. The particle identification method is different between low- and high-$p_{T}$ regions. The main particle detector at STAR, the time projection chamber (TPC), is used to obtain momentum of charged particles in low-$p_{T}$ region of $0.4<p_{T}<0.8$ GeV/c~\cite{STAREfficiency}. And the time-of-flight (TOF) detector is used to do the particle identification in the relatively-high-$p_{T}$ region of $0.8<p_{T}<2$ GeV/c~\cite{STARMoment,LuoCPOD}. In this case, particles need to be counted separately for the two $p_{T}$ regions, in which the values of the efficiencies are different. 

%%%%%%%%%%%%%%%%%%%%%%%%Figure TPC Efficiency
\begin{figure}[!htp]
\hspace{-0.8cm}
     \includegraphics[scale=0.45]{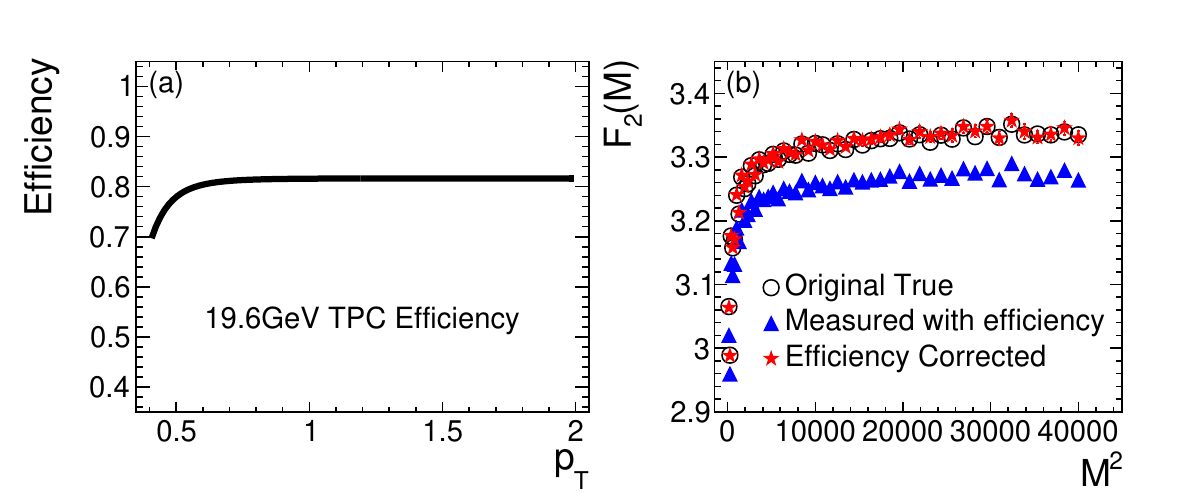}
     
      \caption{(a) Experimental tracking efficiencies as a function of $p_{T}$ in TPC detector at midrapidity ($|y|<0.5$) for protons in $0\%-5\%$ Au $+$ Au collisions. (b) The second-order SFM as a function of number of partitioned cells from UrQMD calculations. The black circles represent the original true $F_{2}(M)$, the blue solid triangles are the measured $F_{2}(M)$ after discarding particles according to the TPC efficiency, and the red stars show the efficiency corrected SFMs by using the cell-by-cell method.}
 \label{Fig:F2TPC}
 \end{figure} 

In Fig. 4(a), we show the $p_{T}$ dependence of the experimental efficiency in only the TPC detector in the midrapidity ($|y|<0.5$) region for protons in the most central Au $+$ Au collisions at $\sqrt{s_\mathrm{NN}}$ = 19.6 GeV~\cite{STAREfficiency}. It first increases with increasing $p_{T}$, and then gets saturated in higher-$p_{T}$ regions. We employ this tracking efficiency into the UrQMD event sample by keeping a particle according to the probability reading from Fig. 4(a) with the $p_{T}$ of that particle. And the measured $F_{2}(M)$ is calculated in the event sample after discarding particles. Next, we apply the correction formula of Eq.~\eqref{Eq:FMcorrection} to do the efficiency correction on the measured $F_{2}(M)$. It is observed in Fig. 4(b) that the measured SFMs (blue triangle) are systematically smaller than the original true ones (black circles), especially in the large number of partitioned cells. However, the efficiency corrected SFMs (red stars) are found to be well consistent with the original true ones. 

%%%%%%%%%%%%%%%%%%%%%%%%Figure TPC+TOF Efficiency
\begin{figure}[hbtp]
\hspace{-0.8cm}
 \includegraphics[scale=0.45]{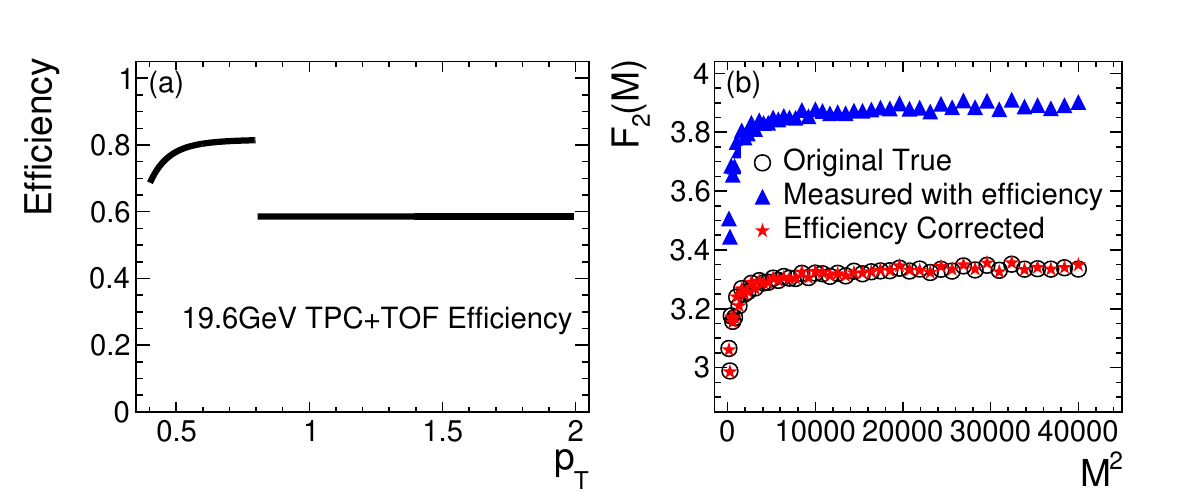}
  \caption{(a) Experimental tracking efficiencies as a function of $p_{T}$ in TPC + TOF detectors at midrapidity ($|y|<0.5$) for protons in $0\%-5\%$ Au $+$ Au collisions. (b) The second-order SFM as a function of number of partitioned cells from UrQMD calculations. The black circles represent the original true $F_{2}(M)$, the blue solid triangles are the measured $F_{2}(M)$ after discarding particles according to the TPC+TOF efficiency, and the red stars show the efficiency corrected SFMs by using the cell-by-cell method.}
    \label{Fig:F2TPTOF}
\end{figure}   

For the case of TPC + TOF efficiencies, Fig. 5(a) shows the tracking efficiencies as a function of $p_{T}$ in TPC and TOF at STAR~\cite{EfficiencyToshihiro,STARMoment,LuoCPOD,EfficiencyTrack}. One notes that there is a steplike dependence of the efficiencies on $p_{T}$. The reason is that the particle identification method is different between TPC and TOF detectors in the STAR experiment. We apply the TPC + TOF efficiency effect to the UrQMD event sample at $\sqrt{s_\mathrm{NN}}$ = 19.6 GeV and then correct the measured SFMs by Eq.~\eqref{Eq:FMcorrection}. The results are shown in Fig. 5(b). Again, the SFMs corrected by the proposed cell-by-cell method (red stars) are verified to be coincide with the original true ones (black circles). 

In this section, we have demonstrated that the cell-by-cell method could serve as a precise and effective way of efficiency correction of SFMs. It can be easily applied to current studies at STAR~\cite{STARintermittency}, NA49~\cite{NA49SFM}, NA61~\cite{NA61universe} and other heavy-ion experiments in the intermittency analysis. It should also be noted that one needs to consider how to treat the momentum resolution in different experiments. Since we use the $p_{T}$ of individual particles to get the efficiency, the momentum resolution might directly affect the calculation of SFMs. This effect could be studied by smearing the $p_{T}$ for each particle with the known value of the momentum resolution.

%%%%%%%%%%%%%%%%%%%%%%%%%%%%%%%%%%%%%%%%%%%%%%%%%%%%%%%%%%%%%%%%%%%%%%%%%%%%%%%
\section{summary and outlook}

In summary, we investigate collision energy and centrality dependence of the SFMs in Au $+$ Au collisions at $\sqrt{s_\mathrm{NN}}=7.7-200$ GeV by using the UrQMD model. The second-order intermittency index is found to be small but nonzero in the transport model without implementing any critical related self-similar fluctuations. A cumulative variable method is then proposed to remove background contributions in the intermittency analysis. It has been verified that this method can effectively reduce the distortion of a Gaussian background from a pure self-similar event sample generated by the CMC model. After applying the method to the UrQMD event sample, we find that the noncritical background effect can be efficiently removed and that value of the intermittency index is close to zero.

In the experimental measurements of intermittency, the measured SFMs should be corrected for detecting efficiencies. We derive a cell-by-cell formula in the calculation of SFMs in heavy-ion collisions. The validity of the method has been checked with the UrQMD model by embedding the tracking efficiencies used in the RHIC (STAR) experiment. It is demonstrated that the cell-by-cell method provides a precise and effective way for the efficiency correction on SFMs. The correction method is universal and can be applied to the ongoing studies of intermittency in heavy-ion experiments. 

%%%%%%%%%%%%%%%%%Figure summary
 \begin{figure} 
     \includegraphics[scale=0.35]{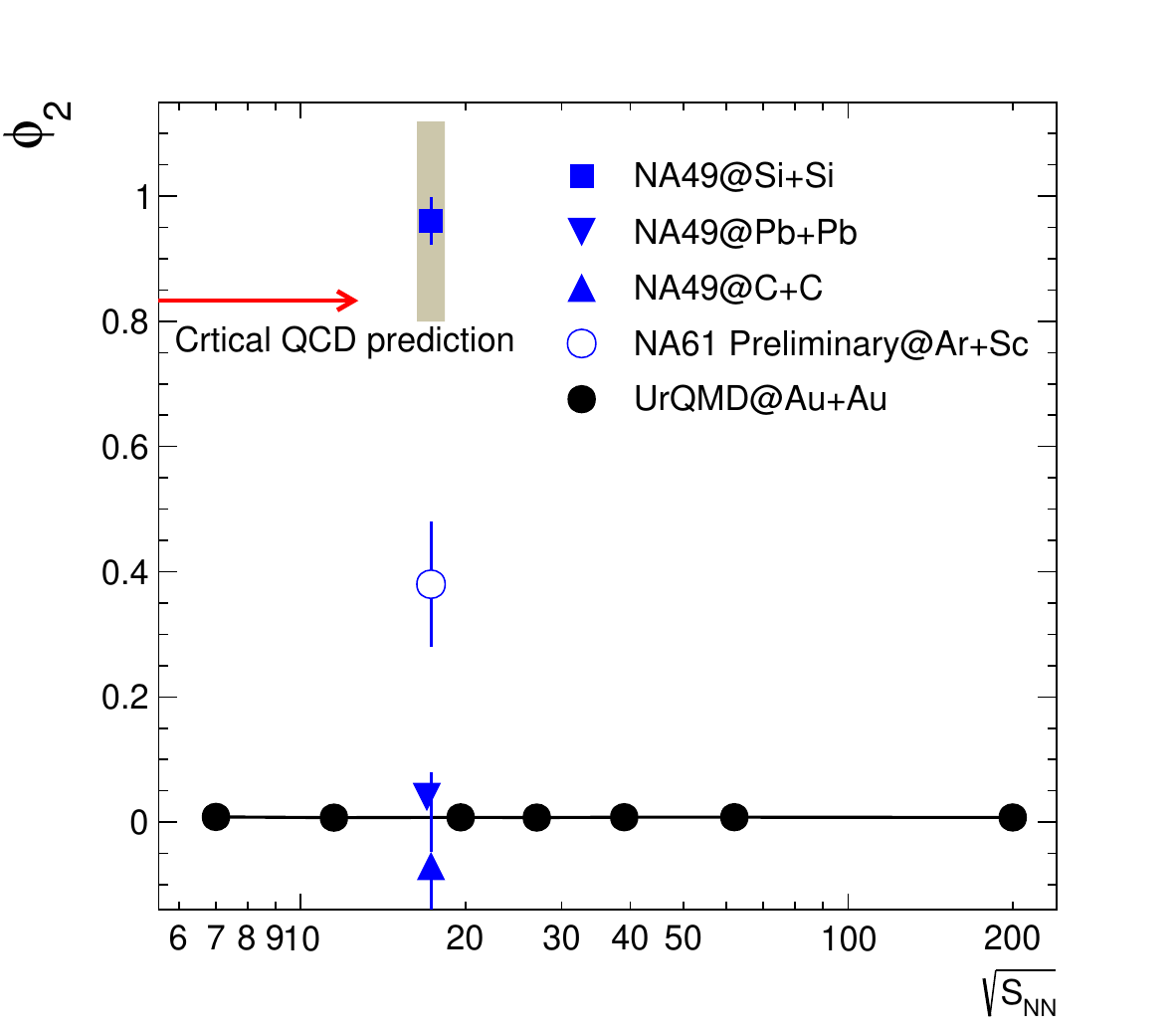}
     \caption{The second-order intermittency index measured at NA49~\cite{NA49SFM,NA49C+C} (solid blue symbols) and NA61~\cite{NA61Ar+Sc} (open blue circles). The results from the UrQMD model in central Au $+$ Au collisions are plotted as black circles. The red arrow represents the theoretic expectation from a critical QCD model~\cite{Antoniou2006PRL}.}
     \label{Figphi2-energy}
   \end{figure}   
%%%%%%%%%%%%%%%%%Figure summary

In current experimental explorations of the intermittency in heavy-ion collisions, the NA49 and NA61 collaborations have directly measured $\phi_{2}$ at various sizes of colliding nuclei~\cite{NA49SFM,NA49C+C,NA61Ar+Sc}, which are represented as blue symbols in Fig. 6. The intermittency parameter at $\sqrt{s_\mathrm{NN}}$ = 17.3 GeV for the Si $+$ Si system at NA49 experiment approaches the theoretic expectation value, shown as red arrow in the figure, in the second-order phase transition in a critical QCD model~\cite{Antoniou2006PRL}. The black circles of the UrQMD results give a flat trend with the value around zero at all energies because no critical mechanisms are implemented in the transport model.

The RHIC (STAR) experiment has finished taking the second phase of beam energy scan (BES-II) program in 2018-2021~\cite{BESII}. With significant improved statistics and detector upgrades in BES-II, it would be interesting that the STAR experiment could measure intermittency to explore the CP in the QCD phase diagram. Our work provides a noncritical baseline and gives a guidance of background subtraction and efficiency correction for the calculations of intermittency in heavy-ion collisions.  

%%%%%%%%%%%%%%%%%%%%%%%%%%%%%%%%%%%%%%%%%%%%%%%%%%%%%%%%%%%%%%%%%%%%%%%%%%%%%%%
\section* {Acknowledgements}
This work is supported by the National Key Research and Development Program of China (Grants No. 2020YFE0202002 and No. 2018YFE0205201), the National Natural Science Foundation of China (Grants No. 12122505, No. 11890711 and No. 11861131009).  The Innovation Fund of the Key Laboratory of Quark and Leptonic Physics (Grant No. QLPL2020P01), and the Ministry of Science and Technology (MoST) under grant No. 2016YFE0104800 are also acknowledged. 

J.W. and Y.L. contributed equally to this work.
%%%%%%%%%%%%%%%%%%%%%%%%%%%%%%%%%%%%%%%%%%%%%%%%%%%%%%%%%%%%%%%%%%%%%%%%%%%%%%%
%\bibliography{ref}% BiTex form

\begin{thebibliography}{99}
%------------------------Paragraph 1
\bibitem{StephanovPD} M. A. Stephanov, K. Rajagopal, E. V. Shuryak, Phys. Rev. Lett. 81, 4816 (1998).
\bibitem{MomentOverview} X. Luo, N. Xu, Nucl. Sci. Technol. 28, 112 (2017).
\bibitem{QCDReport} A. Bzdak, S. Esumi, V. Koch, J. Liao, M. Stephanov, N. Xu, Phys. Rept. 853, 1-87 (2020).
\bibitem{conservecharge0} M. Asakawa, U. W. Heinz, B. Muller, Phys. Rev. Lett. 85, 2072 (2000).
\bibitem{conservecharge1} V. Koch, A. Majumder, J. Randrup, Phys. Rev. Lett. 95, 182301 (2005).
\bibitem{CEP1}M. Stephanov, Prog. Theor. Phys. Suppl. 153, 139 (2004); Z. Fodor and S. D. Katz, J. High Energy Phys. 04, 050 (2004).
\bibitem{CEP2} R. V. Gavai, S. Gupta, Phys. Rev. D 78, 114503 (2008); Phys. Rev. D 71, 114014 (2005).
\bibitem{CEP3} W. Fu, X. Luo, J. M. Pawlowski, F. Rennecke, R. Wen, S. Yin, arXiv: 2101.06035 [hep-ph] (2021). 
\bibitem{searchCEP1} M. A. Stephanov, Phys. Rev. Lett. 102, 032301 (2009).
\bibitem{searchCEP2} M. A. Stephanov, Phys. Rev. Lett. 107, 052301 (2011).
\bibitem{searchCEP3} B. J. Schaefer, M. Wagner, Phys. Rev. D 85, 034027 (2012).
\bibitem{searchCEP4} S. Gupta, X. Luo, B. Mohanty, H. G. Ritter, N. Xu, Science 332, 1525 (2011).
\bibitem{STARPRLMoment} J. Adam {\it et al.} (STAR Collaboration), Phys. Rev. Lett. 126, 092301 (2021)
\bibitem{invariant1} J. Wosiek, Acta Phys. Polon. B 19, 863 (1988).
\bibitem{invariant2} H. Satz, Nucl. Phys. B 326, 613 (1989).
\bibitem{invariant3} E. A. De Wolf, I. M. Dremin, W. Kittel, Phys. Rept. 270, 1 (1996).
\bibitem{Antoniou2006PRL} N. G. Antoniou, F. K. Diakonos, A. S. Kapoyannis, Phys. Rev. Lett. 97, 032002 (2006).
\bibitem{Antoniou2010PRC} N. G. Antoniou, F. K. Diakonos, A. S. Kapoyannis, Phys. Rev. C 81, 011901 (2010).
\bibitem{Antoniou2016PRC} N. G. Antoniou, N. Davis, F. K. Diakonos, Phys. Rev. C 93, 014908 (2016).
\bibitem{Antoniou2018PRD} N. G. Antoniou, F. K. Diakonos, X. N. Maintas, C. E. Tsagkarakis, Phys. Rev. D 97, 034015 (2018).
\bibitem{NA49SFM} T. Anticic {\it et al.} (NA49 Collaboration), Eur. Phys. J. C 75, 587 (2015).
\bibitem{NA61universe} D. Prokhorova and N. Davis (for the NA61/SHINE Collaboration), Universe 5, 103 (2019).
\bibitem{STARintermittency} J. Wu {\it et al.} (for the STAR Collaboration), presentation at ISMD2021, virtual conference, 12-16 July 2021.
\bibitem{CMCPLB} J. Wu, Y. F. Lin, Y. F. Wu, Z. M. Li, Phys. Lett. B 801, 135186 (2020).
\bibitem{UrQMDLi} P. C. Li, Y. J. Wang, J. Steinheimer, Q. F. Li, H. F. Zhang, Phys. Lett. B 818, 136393 (2021).
\bibitem{IJMPE} S. Ahmad, M. M. Khan, S. Khan, Int. J of Mod. Phys. E 23, 1450065 (2014).
\bibitem{amptNPA} X. Y. Long, C. Gang, W. J. Ling {\it et al.}, Nucl. Phys. A 920, 33 (2013).
\bibitem{epos} R. Gupta and S. K. Malik, Adv. High Energy Phys. 2020, 5073042 (2020).
\bibitem{densityF} V. A. Abramovskii, V. N. Gribov, and O. V. Kancheli, Yad. Fiz. 18, 595 (1973). 
\bibitem{SFM1} A. Bialas, R. Peschanski, Nucl. Phys. B 273, 703 (1986).
\bibitem{APPB} K. Fialkowski, B. Wosiek, J. Wosiek, ACTA Phys. Polo. B 20, 639 (1989).
\bibitem{OchsCumulative} W. Ochs, Z. Phys. C: Part. Fields 50, 339 (1991).
\bibitem{BialasCumulative} A. Bialas, M. Gazdzicki, Phys. Lett. B 252, 483 (1990).
\bibitem{STAREfficiency} L. Adamczyk {\it et al.} (for the STAR Collaboration), Phys. Rev. C 96, 044904 (2017).
\bibitem{EfficiencyLuo} X. Luo, Phys. Rev. C 91, 034907 (2015).
\bibitem{MBUrQMD} M. Bleicher {\it et al.}, J. Phys. G: Nul. Part. Phys. 25, 1859 (1999).
%---------------------------------------------------------------------------

%  Section UrQMD-----------------------------------------------------------------
\bibitem{SABassUrQMD} S. A. Bass {\it et al.}, Prog. Part. Nucl. Phys. 41, 255 (1998).
\bibitem{HPUrQMD} H. Petersen, J. Streinheimer, G. Burau, M. Bleicher, H. Stocker, Phys. Rev. C 78, 044901 (2008).
%--------------------------------------------------------------

%Section, method of analysis--------------
\bibitem{HattaPRL} Y. Hatta,  M. A. Stephanov, Phys. Rev. Lett. 91, 102003 (2003).
\bibitem{NGNPA2005} N. G. Antoniou, Y. F. Contoyiannis, F. K. Diakonos. Nucl. Phys. A 761, 159 (2005).
\bibitem{SFM2} A. Bialas, R. Peschanski, Nucl. Phys. B 308, 857 (1988).

% Section energy and centrality dependence of F2(M)--------------------------
\bibitem{net_proton2014} L. Adamczyk {\it et al.} (STAR Collaboration), Phys. Rev. Lett. 112, 032302 (2014).
\bibitem{BEboostrap} B. Efron and R. Tibshirani, Stat. Sci. 1, 54 (1986).
\bibitem{NA49C+C} T. Anticic {\it et al.} (NA49 Collaboration), Phys. Rev. C 81, 064907 (2010).
\bibitem{pythia} P. Sarma, B. Bhattacharjee, Phys. Rev. C 99, 034901 (2019).
\bibitem{NA61Ar+Sc} N. G. Antoniou, N. Davis, F. K. Diakonos et al., Nucl. Phys. A 1003, 122018 (2020).
% Section cumulative aviable.-----------------
\bibitem{resolution} S. Samanta, T. Czopowicz, and M. Gazdzicki, arXiv:2105.01763.
\bibitem{na22} N. Agababyan {\it et al.} (EHS/NA22 Collaboration), Z. Phys. C 59, 405 (1993).
% Section efficiency correction
\bibitem{embedding} B. I. Abelev {\it et al.} (STAR Collaboration), Phys. Rev. C 79, 034909 (2009).
\bibitem{EfficiencyMasakiyo} M. Kitazawa, M. Asakawa, Phys. Rev. C 86, 024904 (2012).
\bibitem{EfficiencyPremomoy} P. Ghosh, Phys. Rev. D 85, 054017 (2012).
\bibitem{EfficiencyAdam} A. Bzdak, Volker Koch, Phys. Rev. C 86, 044904 (2012).
\bibitem{EfficiencyKoch} V. Vovchenko, V. Koch, Nucl. Phys. A 1010, 122179 (2021).
\bibitem{EfficiencyHeshu} S. He, X. Luo, Chin. Phys. C 42, 104001 (2018).
\bibitem{momentHADES} J. Adamczwwski-Musch {\it et al.} (HADES Collaboration), Phys. Rev. C 102, 024914 (2020).
\bibitem{STARMoment} M. S. Abdallah {\it et al.} (STAR Collaboration), Phys. Rev. C 104, 024902 (2021).
\bibitem{EfficiencyFM} A. Bzdak, V. Koch, Phys. Rev. C 91, 027901 (2015).
\bibitem{EfficiencyToshihiro} T. Nonaka, M. Kitazawa, S. Esumi, Phys. Rev. C 95, 064912 (2017).
\bibitem{LuoCPOD} X. Luo (for the STAR Collaboration), in Proceedings of the 9th International Workshop on Critical Point and Onset of Deconfinement (CPOD 2014), Bielefeld, Germany, November 17–21, 2014 [PoS CPOD2014, 019 (2015)]. 
\bibitem{EfficiencyTrack} X. Luo, T. Nonaka, Phys. Rev. C 99, 044917 (2019).
% Summary------------------------
\bibitem{BESII} BES-II White Paper (STAR Note): https://drupal.star.bnl.gov/STAR/starnotes/public/sn0598.
\end{thebibliography}
%\include{ref}
%%%%%%%%%%%%%%%%%% Citation %%%%%%%%%%%%%%%%%%%%%%%%%
%\clearpage

\end{document}